\documentclass[aps,pra,twocolumn,superscriptaddress,showpacs,showkeys]{revtex4}
\usepackage[dvips]{graphicx}
\usepackage{bm}
\def\ie{{\it i.e.,\/}}                     
\def\eg{{\it e.g.,\/}}                     
\def\vs{{\it vs.\/}}                       
\def\via{{\it via\/}}                      
\def\cf{{\it cf.\/}}                       
\def\etal{{\it et al.\/}}                  
\def\Na{$\rm Na_2$}                        
\def\waveno{\rm cm^{-1}}                   
\def\Xsig  {${\rm X}^1\Sigma_g^+$}         
\def\Asig  {${\rm A}^1\Sigma_u^+$}         
\def\Bpi   {${\rm B}^1\Pi_u$}              
\def\bpi   {${\rm b}^3\Pi_u$}              
\def\bpij  {${\rm b}^3\Pi_{\Omega u}$}     
\def\bpix#1{${\rm b}^3\Pi_{\Omega=#1,u}$}  

\begin{document}

\title{Laser-induced atomic fragment fluorescence spectroscopy:\\A facile technique for molecular spectroscopy of spin-forbidden states}


\author{ Qun Zhang }
        \thanks{Corresponding authors}
        \email{qunzh@ustc.edu.cn}
        \affiliation{Hefei National Laboratory for Physical Sciences at the Microscale, University of Science and Technology of China, Hefei, Anhui 230026, People's Republic of China}
                \affiliation{Department of Chemical Physics, University of Science and Technology of China, Hefei, Anhui 230026, People's Republic of China}
\author{ Yang Chen }
        \affiliation{Hefei National Laboratory for Physical Sciences at the Microscale, University of Science and Technology of China, Hefei, Anhui 230026, People's Republic of China}
                \affiliation{Department of Chemical Physics, University of Science and Technology of China, Hefei, Anhui 230026, People's Republic of China}
\author{ Mark Keil }
  \thanks{Corresponding authors}
        \email{keil@bgu.ac.il}
        \affiliation{Department of Physics, Ben-Gurion University, Be'er Sheva 84105, Israel}

\date{\today}

\pacs{39.30.+w, 33.90.+h, 32.50.+d, 33.20.Vq, 33.20.Wr}

\keywords{fluorescence spectroscopy, spin-forbidden molecular transitions, perturbation, photodissociation, sodium molecule, laser-induced fluorescence, atomic fragments}

\begin{abstract}

Spectra of spin-forbidden and spin-allowed transitions in the mixed~\bpi$\sim$\Asig\ state of~\Na\ 
are measured separately by two-photon excitation using a single tunable dye laser. The two-photon excitation produces~$\rm Na^*(3p)$ by photodissociation, which is easily and sensitively detected by atomic fluorescence. At low laser power, only the~\Asig\ state is excited, completely free of triplet excitation. At high laser power, photodissociation \via\ the~\bpi\ triplet state intermediate becomes much more likely, effectively ``switching'' the observations from singlet spectroscopy to triplet spectroscopy with only minor apparatus changes. This technique of perturbation-assisted laser-induced atomic fragment fluorescence~(LIAFF) may therefore be especially useful as a general vehicle for investigating perturbation-related physics pertinent to the spin-forbidden states, as well as for studying allowed and forbidden states of other molecules.

\end{abstract}

\maketitle


\section{Introduction}

The well-known \bpi$\sim$\Asig--\Xsig\ inter-combination band system of the~\Na\ molecule has long served as a prototype for developing new spectroscopic techniques~\cite{kaminsky,lyyra,kramer} and for rigorous spectroscopic analysis~\cite{qi}. Although transitions to the lowest-lying triplet~\bpi\ state are spin-forbidden, this state is perturbed by the energetically nearby spin-allowed~\Asig\ state and was first observed almost a century ago~\cite{wood,fredrickson,carroll}. Kusch and Hessel~\cite{kusch} used direct absorption spectroscopy to characterize~\Asig\ state perturbations caused by the~\bpij\ state, with~$\Omega=0,1,2$ representing the projections of electronic angular momenta along the molecular axis, according to Hund's case~(a)~\cite{herzberg}.

Despite being partially allowed due to these perturbations, fully resolving the three~$\Omega$ components of the~\bpij\ state is by no means a trivial task. As estimated by Kr\"amer \etal~\cite{kramer}, the Franck-Condon~(FC) factor for the strongest vibrational band ($v^b=0\leftarrow v''=0$) of the ``forbidden'' \bpix{0}$\leftarrow$\Xsig\ transition is at least two orders of magnitude smaller than the corresponding~$v^A=0\leftarrow v''=0$ band of the ``allowed'' \Asig$\leftarrow$\Xsig\ transition, and~FC factors for the~$\Omega=1,2$ components are respectively~20 and~60 times smaller yet~\cite{kramer}. The absorption intensity of the higher vibrational bands falls strongly with increasing~$v^b$ which makes them even harder to detect.

The inaccessibility of the spin-forbidden~\bpi\ state via traditional spectroscopic methods has engendered many modern techniques since 1975~\cite{kaminsky,lyyra,kramer,%
engelke,atkinson1,atkinson2,shimizu1,shimizu2,li1,effantin,li2,kato1,kato2,whang}, amongst which the most recent and sophisticated are perturbation-facilitated all-optical triple resonance~(PFAOTR) spectroscopy~\cite{whang} and high-resolution resonant two-photon ionization spectroscopy~\cite{kramer}. To date, the three~$\Omega$ components of the~\bpij\ state have been fully resolved only by the continuous wave~(cw)~PFAOTR technique which demands three~cw lasers, all of which must be ``locked'' on the transitions~\cite{whang}.

We report here a newly developed and simple technique of perturbation-assisted laser-induced atomic fragment fluorescence~(LIAFF) spectroscopy that is suitable for detailed studies of spin-forbidden electronic states, and in particular, we have used the perturbation-assisted~LIAFF technique to fully resolve all three~$\Omega$ components of the~\bpij\ state in~\Na. Based on a single conventional pulsed dye laser, the~LIAFF technique does not attain the high resolution of the sophisticated techniques mentioned above; instead we focus here on demonstrating its robustness and experimental utility for easily obtaining information on spin-forbidden states. We expect that the technique will be widely applicable, as shown by our example studies herein on the \bpi$\sim$\Asig--\Xsig\ inter-combination band system of the~\Na\ molecule.

\section{Experimental}

\begin{figure}[h]
  \includegraphics[width=0.45\textwidth]{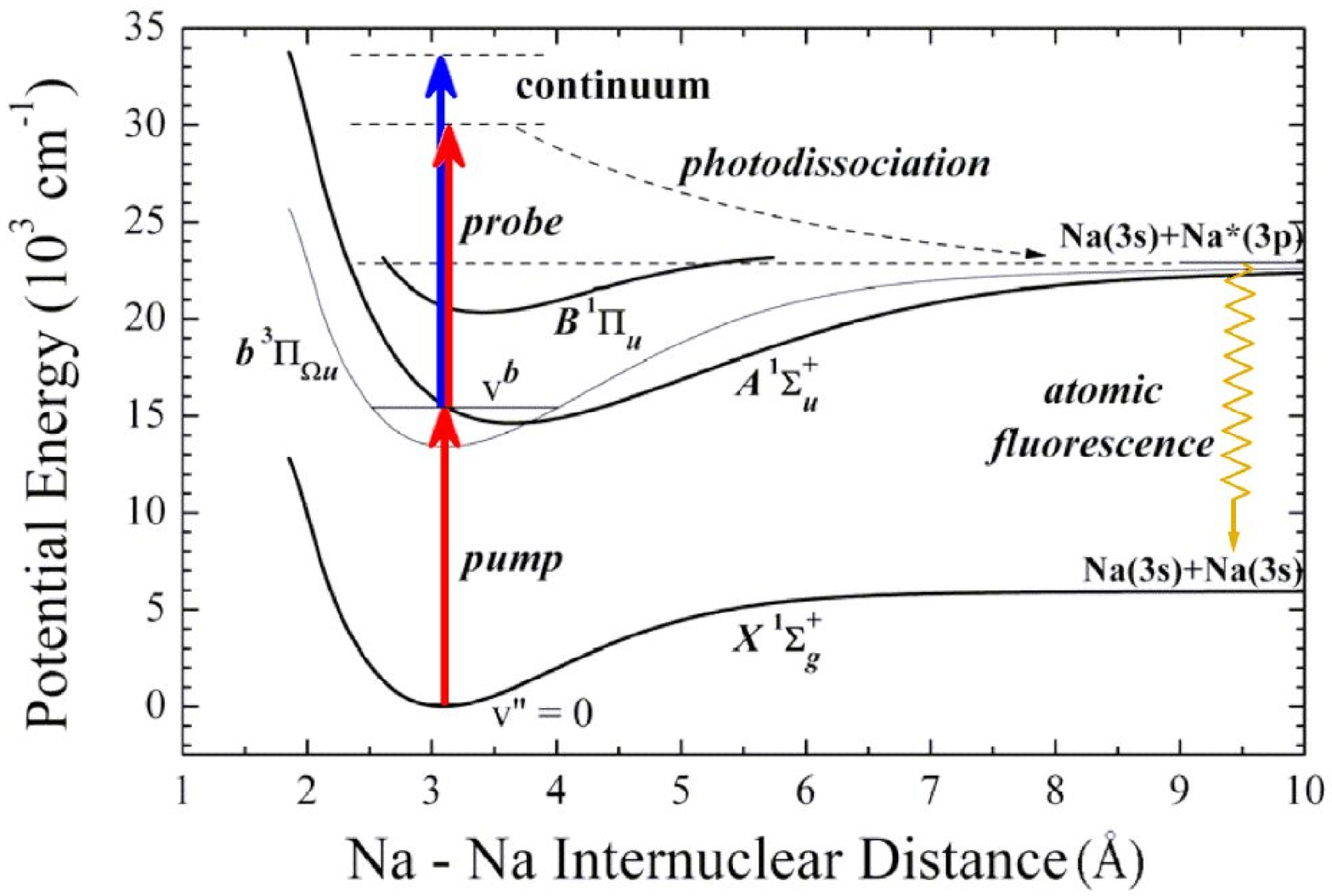}\\
  \caption{(Color online) Electronic energy levels for~\Na. The probe laser may reach the dissociation continuum using the same frequency as the pump laser (short red arrow -- no~YAG mixing in Fig.~\ref{fig:LIAFF_schematic}), or it may have a higher frequency (long blue arrow). In both cases, excitation proceeds \via\ the intermediate~\bpij$\sim$\Asig\ mixed state. Two-photon absorption is energetically required to the reach continuum states photodissociating to excited~$\rm Na^*(3p)$; we detect the resulting atomic fluorescence (wavy yellow arrow.)}
\label{fig:LIAFF_excitation}
\end{figure}

The perturbation-assisted~LIAFF technique is illustrated schematically in Fig.~\ref{fig:LIAFF_excitation} by referring to the four lowest electronic states of~\Na. A pump laser is used to excite ground-state molecules~(\Xsig,$v''=0$) to the spin-forbidden state~(\bpij,$v^b$) that is perturbed by the nearby spin-allowed state~(\Asig). The same laser also
acts as a probe by exciting the mixed~\bpij$\sim$\Asig\ levels to the continuum, dissociating the molecule into $\rm Na(3s)+Na^*(3p)$ atomic fragments. The~$\rm Na^*(3p)$ then decays spontaneously, giving rise to very intense~D-line fluorescence at~$\rm589.6\,nm$~\cite{NIST}. Spectra are obtained explicitly by measuring the atomic fluorescence intensity as a function of the pump laser frequency.  Since detecting fluorescence from a single atomic state is far more efficient than collecting fluorescence from myriad vibrational-rotational molecular states, and can be made very selective by simple optical filters, the perturbation-assisted~LIAFF spectroscopy provides much higher sensitivity for directly and fully observing a spin-forbidden state. Such high sensitivity usually is not attainable by conventional methods (\eg\ molecular laser-induced fluorescence spectroscopy\footnote{Laser-induced fluorescence spectra of the~\bpij--\Xsig\ transition in~\Na\ have been measured for~$\Omega=0,1$ in Ref.~\cite{shimizu2} by modulating the laser and then gating the photon counter, but these authors did not measure the~$\Omega=2$ component.}), or it requires more sophisticated techniques such as the three-cw-laser PFAOTR spectroscopy~\cite{kramer} mentioned above.  The present technique requires only a single conventional tunable pulsed dye laser.

\begin{figure}[h]
  \includegraphics[width=0.45\textwidth]{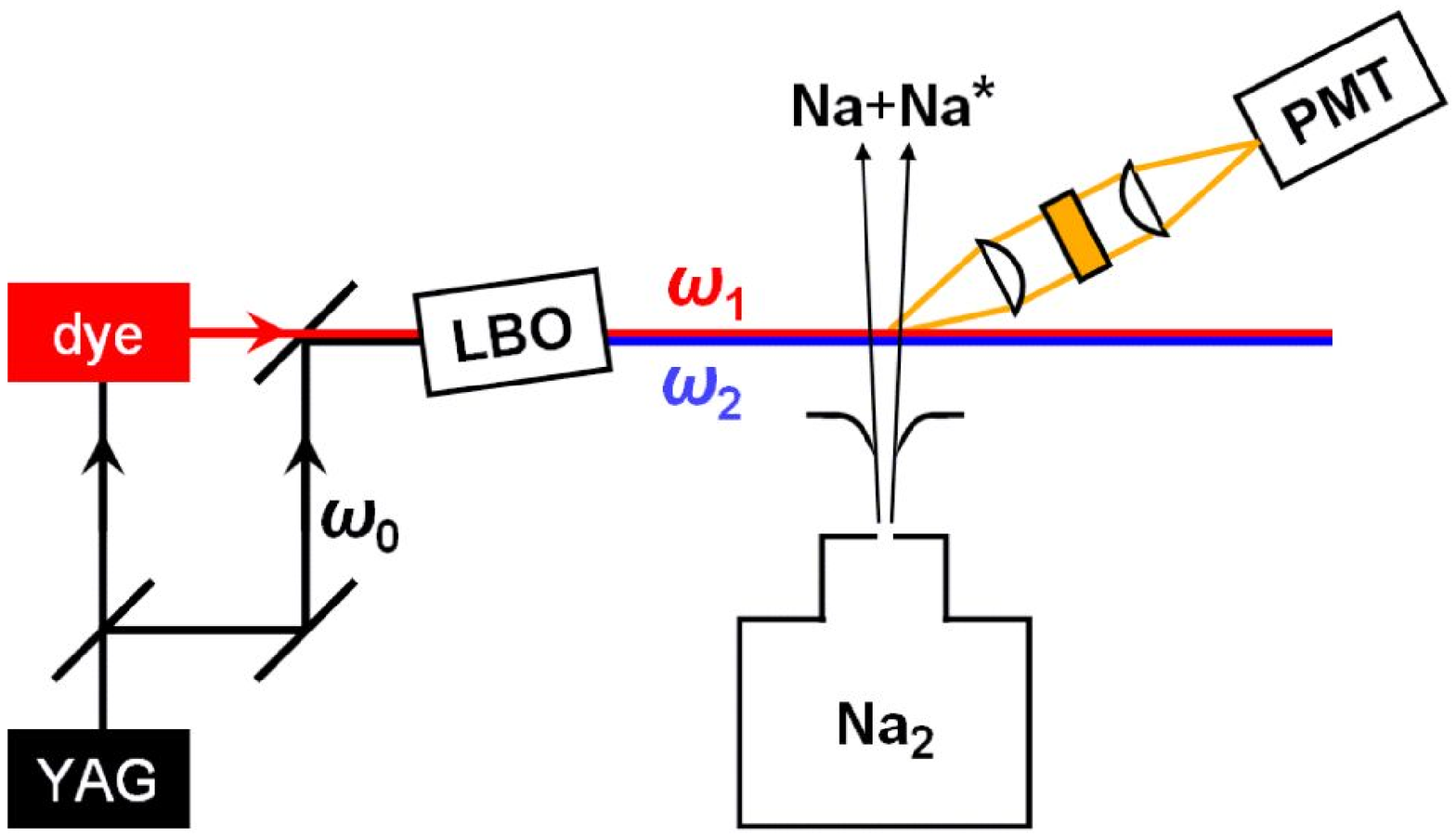}\\
  \caption{(Color online) Sketch of the laser-induced atomic fragment fluorescence~(LIAFF) apparatus. A single dye laser (red laser: frequency~$\omega_1$) is used for two-photon photodissociation of~\Na\ in a molecular beam. The molecular beam expansion ensures that most of the~\Na\ molecules are in the ground vibrational state~\Xsig($v''=0$). The~Nd:YAG laser used to pump the dye laser may also be used to generate a second frequency (blue laser:~$\omega_2=\omega_1+\omega_0$) by non-linear mixing in the angle-tuned lithium tri-borate~(LBO) crystal. Atomic fluorescence from~$\rm Na^*(3p)$ is collimated and observed through a narrow-band interference filter before detection by the photomultiplier tube.}
\label{fig:LIAFF_schematic}
\end{figure}

The experiment is conducted in a supersonic jet environment. A schematic diagram of the apparatus is shown in Fig.~\ref{fig:LIAFF_schematic}. The~\Na\ molecules are formed during the adiabatic expansion of pure sodium vapor from a conventional two-chamber oven heated to~$\rm\sim900\,K$. The beam of cold~\Na\ molecules is directed into a differentially pumped observation chamber through a collimating skimmer, and is then intersected perpendicularly with a tunable dye laser (Continuum TDL-60) pumped by a conventional~10-Hz pulsed~Nd:YAG laser (Continuum 682-10). The dye laser ($\rm5\,ns$~pulse duration) used~DCM and~LDS698 dyes and its linewidth is~$\rm0.1\,\waveno$; wavelengths are calibrated by a pulsed Fabry-Perot wavemeter. The dye laser could easily be adjusted to provide a peak intensity between~$\rm1\,kW/cm^2$ and~$\rm10\,MW/cm^2$; for some experiments the dye output is mixed with the~Nd:YAG output to provide higher probe frequencies with a peak intensity up to~$\rm2\,MW/cm^2$. Laser-induced fluorescence is collected by an~$f/4$ optical collimating lens system and photomultiplier tube. A narrow-band interference filter ($\rm3\,nm$ full-width half-maximum) is placed between the lenses to transmit only the fluorescence of interest, \ie\ the~Na D-line in this proof-of-principle experiment.

\section{Results and discussion}

\begin{figure}[h]
  \includegraphics[width=0.45\textwidth]{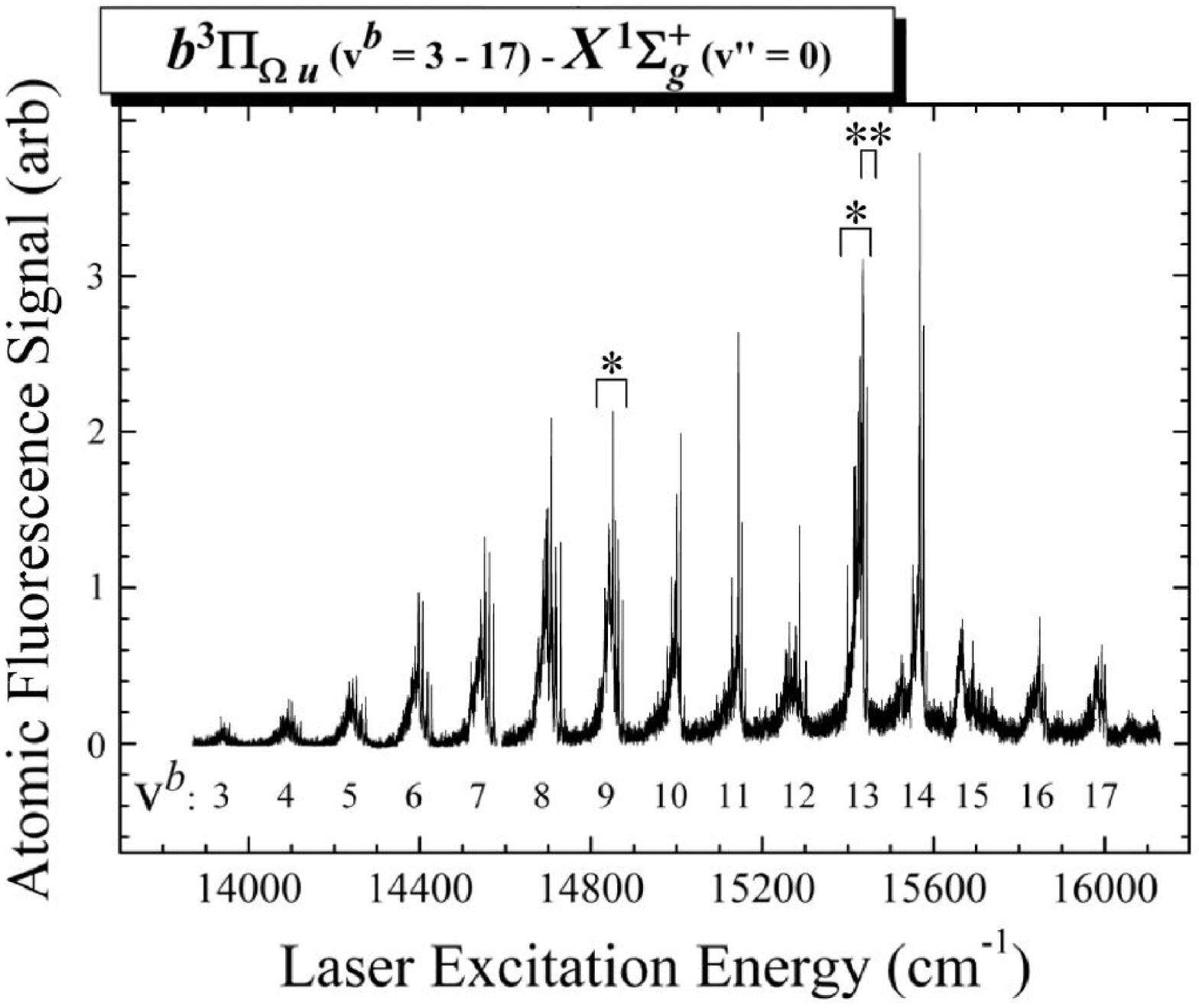}\\
  \caption{Survey spectrum of~\bpij($v^b$)$\leftarrow$\Asig($v''=0$) vibrationally resolved transitions obtained using the perturbation-assisted~LIAFF technique reported here. A single tunable pulsed dye laser, used as both pump and probe, is used to obtain this spectrum. The laser is operated at the relatively high peak intensity of~$\rm1-3\,MW/cm^2$. The two rovibrational bands marked by single asterisks are presented in more detail in Fig.~\ref{fig:LIAFF_omega}, while the narrow region marked with the double-asterisk is expanded in Fig.~\ref{fig:LIAFF_singlet}. The vibrational assignments~\cite{shimizu2,whang} correspond to the single-photon excitation energy~$\hbar\omega_1$ shown; the total excitation energy to the dissociation continuum is~$2\hbar\omega_1$. The vertical axis records the photomultiplier signal in arbitrary units.}
\label{fig:LIAFF_survey}
\end{figure}

\begin{figure}[h]
  \includegraphics[width=0.45\textwidth]{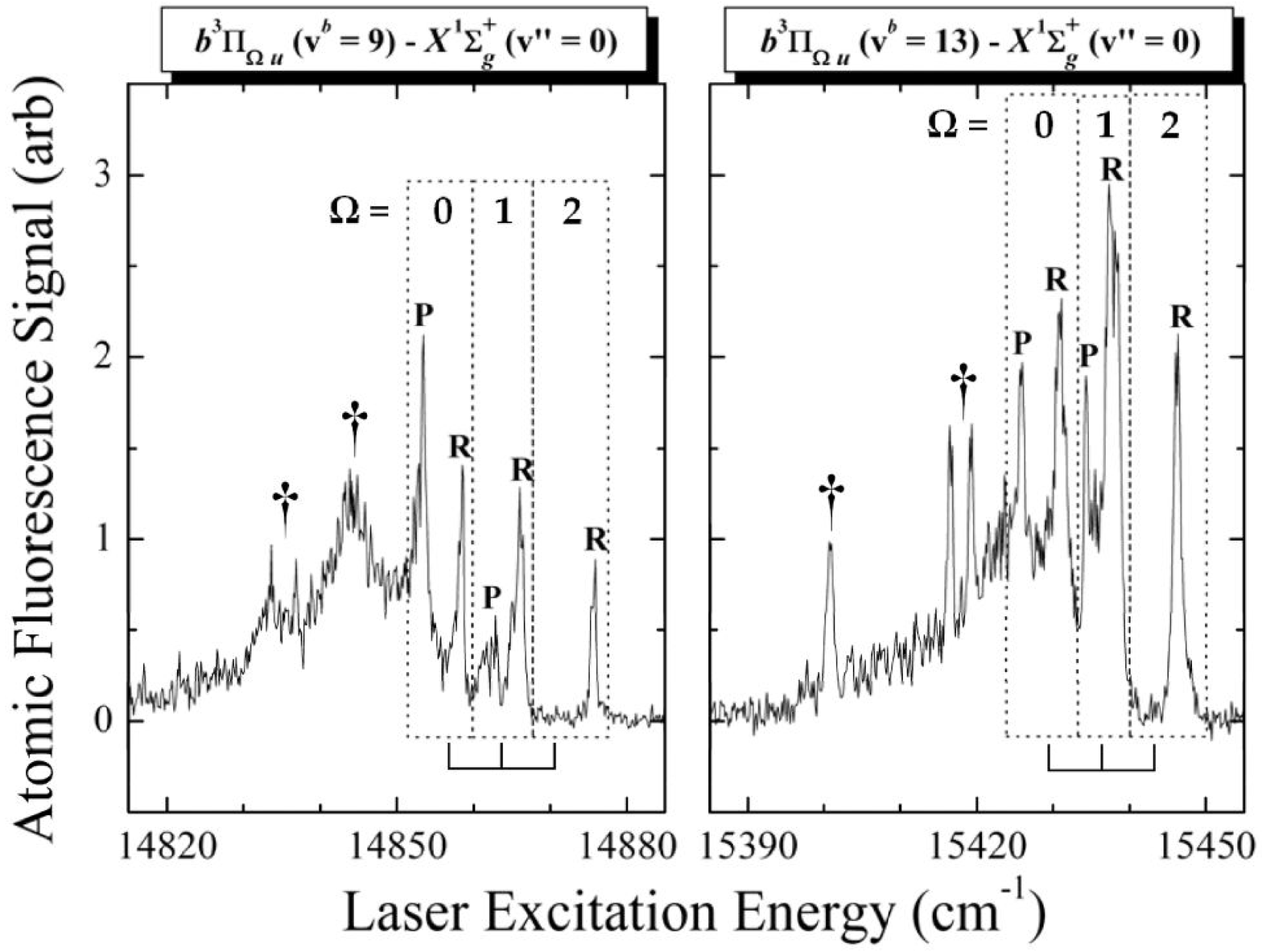}\\
  \caption{Detailed scan of the regions marked by single asterisks in Fig.~\ref{fig:LIAFF_survey}, showing perturbation-assisted~LIAFF spectra for the spin-forbidden~\Na\ \bpix{0,1,2}($v^b=9,13$)$\leftarrow$\Xsig($v''=0$) transitions. Vertical ticks indicate vibrational band origins of the three Hund's case~(a) components~$\Omega=0,1,2$, respectively~\cite{shimizu2,whang}. The energy region for each~$\Omega$ component is indicated by dotted boxes, with~P's and~R's labeling the corresponding branches. The spectral features marked by daggers remain unassigned in this work. Note however, that there are no features attributable to spin-allowed~\Asig$\leftarrow$\Xsig\ transitions. As in Fig.~\ref{fig:LIAFF_survey}, a single tunable pulsed dye laser is used as both pump and probe at a peak intensity of~$\rm1-3\,MW/cm^2$.}
\label{fig:LIAFF_omega}
\end{figure}

The spin-forbidden~\bpij$\leftarrow$\Xsig\ triplet-singlet transition demands a sufficiently intense laser~\cite{kramer} because of its far smaller transition probability compared to the allowed~\Asig$\leftarrow$\Xsig\ singlet-singlet transition. This is easily accomplished by using the~$\omega_1$ dye laser frequency for both the pump and the probe. Vibrationally well-resolved perturbation-assisted~LIAFF spectra are shown in Fig.~\ref{fig:LIAFF_survey}; based on the data reported in Refs.~\cite{shimizu2} and~\cite{whang}, we can unambiguously assign this survey spectrum to the~\bpij($v^b=3-17$)$\leftarrow$\Xsig($v''=0$) transitions of the~\Na\ molecule. In~Fig.~\ref{fig:LIAFF_omega} we expand the~$(9,0)$ and~$(13,0)$ bands, where we also see clearly resolved rotational~P and~R branches for each~$\Omega=0,1,2$ component of the~\bpij\ state (except for the~P branch of~$\Omega=2$), as depicted within the dotted boxes\footnote{Since the transition moment of the~\bpij--\Xsig\ transition is mostly due to mixing with the singlet~\Asig\ state, the~Q branch transitions are much more difficult to observe, as confirmed previously~\cite{shimizu1,shimizu2,kato1,kato2}.}. Despite much lower~FC factors for direct excitation of the~$\Omega=1,2$ components compared to~$\Omega=0$, the perturbation-assisted~LIAFF spectra exhibit comparable intensities for all three components, demonstrating that this technique provides facile access to the spin-forbidden~\bpij\ state.

The~$\Omega=0$ component of the~\bpij\ state can be readily understood in terms of ``intensity borrowing''~\cite{lefebvre-brion} from the nearby~\Asig\ state \via\ direct spin-orbit coupling~($\Delta\Omega=0$), while the~$\Omega=1,2$ components must borrow intensity from the triplet~$\Omega=0$ component \via\ indirect spin-rotation coupling ($\Delta\Omega=1,2$)~\cite{kramer,lefebvre-brion}. This implies that if the~\Asig\ state is the only perturber, the intensities of transitions into the~$\Omega=0,1,2$ components should be much weaker than for the~$\Omega=0$ component. Our observation of anomalous intensities for these two components, as shown in Fig.~\ref{fig:LIAFF_omega}, indicates that there may be another perturbing state in addition to the~\Asig\ state. The most likely candidate is the singlet~\Bpi\ state which lies close to the~\bpij\ state. The~\Bpi\ state correlates asymptotically to the same~$\rm(3s+3p)$ configuration as do the~\Asig\ and~\bpij\ states (see Fig.~\ref{fig:LIAFF_excitation}). 

If the~\Bpi\ state is indeed a ``remote perturber'', the~P and~R branches belonging to the~$\Omega=1$ component may result from spin-orbit interaction (~$\Delta\Omega=1$) with the~\Bpi\ state as well as from spin-rotation interaction ($\Delta\Omega=1$) with the mixed~\Asig$\sim$\bpix{0} levels, as stated in Ref.~\cite{kramer}. As such, interference between these two effects may dictate the intensity pattern of the~P and~R branches, since the transition amplitude phases for~P and~R branches of parallel transitions~($\Delta\Lambda=0$, here~\Asig$\leftarrow$\Xsig) are identical, while those for~P and~R branches of perpendicular transitions~($\Delta\Lambda=\pm1$, here~\Bpi$\leftarrow$\Xsig) are opposite~\cite{lefebvre-brion}. From the observed spectra shown in Fig.~\ref{fig:LIAFF_omega}, we can infer that such interference is constructive for~R branch transitions and destructive for~P branch transitions in both the~$(9,0)$ and~$(13,0)$ vibrational bands of the~\bpix{1}$\leftarrow$\Xsig\ transition.  Although such an observation differs from the weak~$(0,0)$ vibrational band reported previously~\cite{kramer}, both cases can be valid~\cite{lefebvre-brion,klynning}: if such parallel~\vs~perpendicular interference is constructive for~P branch transitions, it will be destructive for~R branch transitions, and {\it vice versa}; whether the interference is constructive or destructive depends on the intrinsic nature of the rovibronic transition under investigation.

The~P and~R branches belonging to the~$\Omega=2$ component may originate from spin-rotation interaction~($\Delta\Omega=1$) with the mixed~\Asig$\sim$\bpix{1}\ levels as well as from spin-rotation interaction~($\Delta\Omega=2$) with the mixed~\Asig$\sim$\bpix{0}\ levels. Similar to the case for the~$\Omega=1$ component, interference between these two perturbation effects are observed to be constructive for the~R branch transitions, while it is notable that the~P branch transitions are observed to be suppressed almost completely.

Apart from the~$\Omega=0,1,2$ components observed for the~\bpij\ state, there remain several unassigned spectral features, including sharp peaks and broad structures (denoted by daggers in Fig.~\ref{fig:LIAFF_omega}) that extend to the lower energy region of both spectra. These features cannot reasonably be interpreted using just the four electronic states depicted in Fig.~\ref{fig:LIAFF_excitation}. Additional interactions involving other ``remote perturber'' states may be needed to explain these features, which would require further investigations that lie outside the focus of this paper. Nevertheless, the rich information unveiled by the perturbation-assisted~LIAFF spectra demonstrates that the technique can provide not only easy access to a spin-forbidden state, but also insight into perturbation-related physics in molecular systems.

\begin{figure}[h]
  \includegraphics[width=0.45\textwidth]{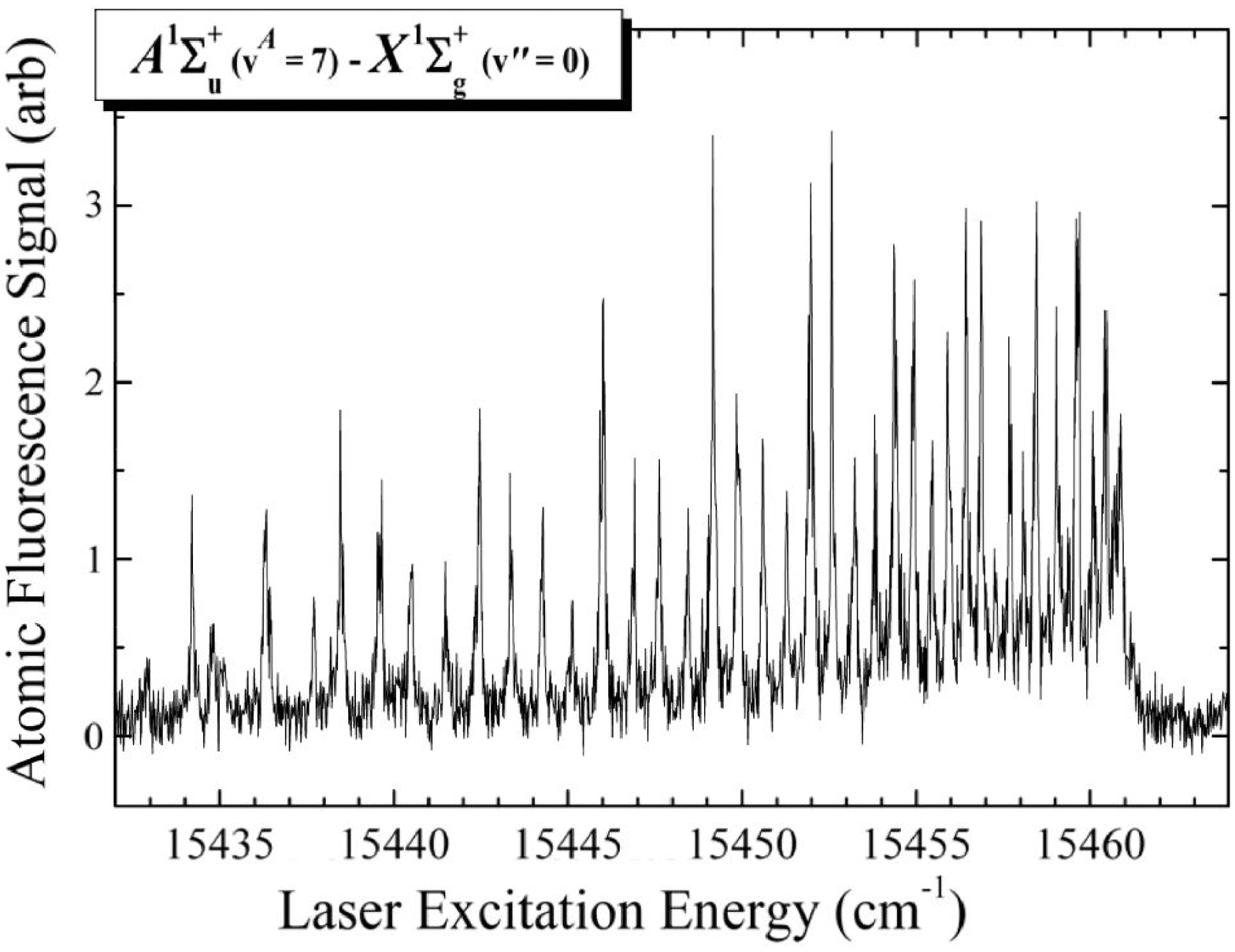}\\
  \caption{Detailed scan of the region marked by a double-asterisk in Fig.~\ref{fig:LIAFF_survey}, showing a rotationally well-resolved spectrum of the spin-allowed \Asig($v^A=7$)$\leftarrow$\Xsig($v''=0$) transition, obtained using the~LIAFF technique. In contrast to Figs.~\ref{fig:LIAFF_survey} and~\ref{fig:LIAFF_omega}, the tunable pump laser is operated at a peak intensity of~$\rm\sim8\,kW/cm^2$, which is too weak by itself to drive two-photon excitation to the continuum. The probe laser is instead provided by mixing residual~Nd:YAG light with the intense remainder of the dye laser light, and has a peak intensity of~$\rm\sim0.5\,MW/cm^2$. The energy axis shown corresponds to the pump laser frequency~$\omega_1$; the total energy corresponds to~$\omega_1+\omega_2=2\omega_1+\omega_0$ (Fig.~\ref{fig:LIAFF_schematic}).}
\label{fig:LIAFF_singlet}
\end{figure}

We now discuss using the~LIAFF technique to ``turn off'' the triplet spectroscopy and replace it with singlet spectroscopy. By lowering the pump laser intensity from~$\rm\sim2\,MW/cm^2$ to~$\rm\sim8\,kW/cm^2$, the ``forbidden'' triplet state spectrum is completely lost because there is no longer enough laser power for the two-photon transitions energetically required for producing electronically excited atomic~Na. The~LIAFF signal drops to zero, even though single-photon 
excitation to the fully ``allowed''~\Asig\ state still occurs (as verified by monitoring {\it molecular} \Asig$\rightarrow$\Xsig\ fluorescence separately). Excitation to the~\Asig\ state in the same spectral region of the~\Na\ molecule can nevertheless be recovered, completely free of triplet excitation: we still use the~LIAFF technique but simply add the higher-frequency~$\omega_2$ probe beam (Fig.~\ref{fig:LIAFF_schematic}). Only the second laser operates at the high intensity ($\rm\sim1\,MW/cm^2$) required for efficient photodissociation. The probe laser is obtained conveniently by mixing the dye laser with the~Nd:YAG fundamental~($\rm1064\,nm$); its wavelength of~$\rm400-410\,nm$ ensures sufficient energy for the atomic fragment fluorescence but it does not yield any such fluorescence without the~$\omega_1$ pump laser. 

Using very low power for the pump laser to avoid power broadening in excess of the laser linewidth, we show a rotationally well-resolved~LIAFF spectrum of the~\eg\ $(7,0)$~vibrational band of the~\Asig$\leftarrow$\Xsig\ transition in Fig.~\ref{fig:LIAFF_singlet}~\cite{herrmann}. It is remarkable that this spectrum of the~\Asig\ state is completely free of triplet features, and yet there is no hint of it in the~\bpi\ spectrum that we measure in exactly the same spectral region (\cf\ overlapping regions marked by the single and double asterisks in Fig.~\ref{fig:LIAFF_survey}) between~15,432 and~15,461$\,\waveno$\footnote{Together with the adjacent~($6,0$) band, individual rotational states in these two bands were selected as the basis for a four-colour two-pathway coherent-control experiment, as discussed in~Refs.~\cite{shapiro,zhang}}, as shown in Fig.~\ref{fig:LIAFF_singlet}. This implies that the~LIAFF technique can be extended to separately and easily study allowed and forbidden states of other molecules if they have detectable fluorescence from their excited atomic fragment(s). 

In the present experiment, laser power broadening precluded full rotational resolution of the~P and~R branches for the spin-forbidden~\bpij\ state. The experimental sensitivity may easily be improved however, \eg\ by using faster optics (we have built an~$f/1$ optical system instead of the~$f/4$ system used in this study) and/or more sensitive and faster fluorescence detection. Such improvements should allow detecting even weaker features in the ``forbidden'' spectrum, or alternatively, improving spectral resolution by lowering the laser power and thereby rotationally resolving \eg\ the~P and~R spectral features with different~$\Omega$ components of the~\bpij\ state.

\section{Conclusion}

We have presented a newly developed technique of perturbation-assisted laser-induced atomic fragment fluorescence~(LIAFF) spectroscopy for probing a spin-forbidden state perturbed by nearby spin-allowed state(s). The robustness of this experimentally facile technique (which requires only a single conventional pulsed dye laser and simple fluorescence collection) has been demonstrated for the \bpij$\sim$\Asig$\leftarrow$\Xsig\ inter-combination band system of the~\Na\ molecule.  The spin-forbidden~\bpij\ state is directly observed and all three fine structure components~($\Omega=0,1,2$) are fully resolved and easily detected in the perturbation-assisted~LIAFF spectra, observations that are much more difficult to achieve using other spectroscopic methods. 
Another intriguing feature of the technique presented in this paper is that it can easily be used to ``switch'' between accurate and sensitive observation of ``allowed'' and ``forbidden'' states with only minor apparatus changes. The perturbation-assisted~LIAFF technique may therefore become a general and simple tool for investigating perturbation-related physics pertinent to spin-forbidden states.  

\begin{acknowledgments}

We gratefully acknowledge Moshe Shapiro (Weizmann Institute) for inspiration, support, and the use of the apparatus for these experiments. Q.~Z. is thankful for support from the National Natural Science Foundation of China (Grant No.~20873133) and the Ministry of Science and Technology of China (Grant No.~2007CB815203). M.~K. thanks the Ministry of Immigrant Absorption (Israel).

\end{acknowledgments}


\end{document}